\documentclass[aps,showpacs,
]{revtex4} 
\usepackage{graphicx}
\begin{document}

\title{\large \bf
On superconducting mechanism in the iron-based layered superconductors}
\author{L. S. Mazov}
\affiliation{Institute for Physics of Microstructures, Russian
Academy of Sciences, Nizhny Novgorod 603600 Russia}
\begin{abstract}
It is demonstrated that superconducting (SC) mechanism in doped
Fe-based compounds is characteristic for itinerant electron
systems with coexistence of both (e-e)- and (e-h)-pairing arising
due to electron-phonon and Coulomb interactions, respectively. In
such systems, with decreasing $T$ first the dielectric (SDW) phase
transition (and structural one) happens at $T = T^*$ (due to
(e-h)-pairing) and only then, at $T = T_c$ ($T_c \le T^*$), the SC
transition starts (due to (e-e)-pairing), so that below $T_c$ the
system enters the coexistence (SC+SDW)-phase. (The thermodynamics
of dielectric (SDW) phase transition here is the same as for
superconductor). The corresponding dielectric (SDW) gap is highly
anisotropic since it is only formed at symmetrical parts of the
Fermi surface, and its magnitude $\Sigma$ is large compared with
that of SC gap $\Delta$ ($\Delta < \Sigma$). The SC transition in
such systems happens at higher temperatures compared with
conventional LTSC in BCS-system. Such high $T_c$ is a natural
consequence of (e-e)-pairing at the background of high density of
states which singularity arises in the narrow energy range near
dielectric-(SDW)-gap edges due to removing of electronic states
from the energy region of dielectric (SDW) gap (already formed in
the normal state). However, formation of the dielectric (SDW) gap
at the Fermi surface leads to decreasing of energy region for
(e-e)-pairing what, in its turn, leads, on the contrary, to
decreasing of $T_c$. So, the maximum appears in the dependence of
$T_c$ (and hence $\Delta$) as a function of doping. In contrast,
the dielectric (SDW) gap $\Sigma$ (and hence $T^*$) is a
decreasing function of doping. Thus, these two doping dependencies
are, in fact, forming a phase diagram of such system with
interplay between of SC and magnetism. These conclusions follow
from detailed analysis of available data on resistivity of doped
Fe-based compounds (with taking into account the magnetic nature
of parent compound ReOFeAs) as well as another experimental data
on the basis of model with partial dielectrization of electron
energy spectra. The picture obtained and manifestation of two
order parameters (SC and SDW) in experiments, first of all, in
threshold phenomena are discussed and some predictions are made.
The comparison with the case of cuprates is performed.
\end{abstract}
\maketitle

\section {INTRODUCTION} There was recently an essential progress
in raise of critical temperature of superconducting transition
$T_c$ in newly discovered iron-based superconductors: for a short
time period the superconducting critical temperature $T_c$ has
been raised from several Kelvins to 55 K \cite{Hosono1, Ch1}.
These superconductors are layered in crystal structure with
alternate stacking $FeAs$ and $LaO$ layers. $FeAs$-plane, with
iron spins ordered antiferromagnetically, is considered as
conducting one (cf. with $CuO_2$-plane in cuprates) while $LaO$
layer serves as reservoir (for charge carriers) under doping.
These compounds become to be superconducting with doping of parent
(non-superconducting) compounds $ReOFeAs$ (where Re is a
rare-earth element) which are a 'bad' (magnetic) metals with
transition to commensurate static spin density wave (SDW) state
below certain temperature ($\sim 150 K$). This SDW transition in
$ReOFeAs$ parent compound is near structural phase transition from
tetragonal to orthorhombic phase what gives rise to anomaly in
resistivity, optical, neutron and other experiments (for review,
see, e.g. \cite{Hosono2}) . However, attempts to detect the
presence of SDW in superconducting Fe-based compounds in neutron,
Mossbauer, transport experiments appear to be mostly unsuccessful,
so that now it is wide belief that under doping of parent
compounds the SDW order in ReOFeAs is displaced by SC one in doped
compounds. In other words, the SDW and SC are considered as
competing orders in these novel superconductors. In present paper,
there are presented results of detailed analysis of resistivity
data as well as another ones for novel iron-based superconductors
on the basis of theory of itinerant electron systems with
interplay between superconductivity and magnetism characterized by
partial dielectrization of the electron energy spectra.

\section {EVIDENCE FOR MAGNETIC (AF SDW) PHASE
TRANSITION BEFORE SC ONE}

As known, the resistivity of magnetic metals and alloys is usually
decomposed in two main contributions: phonon $\rho_{ph}(T)$ and
magnetic $\rho_m(T)$ ones (in neglecting of residual one)(see,
e.g. \cite{Vons})
\begin{figure}
\includegraphics[width=10.5 cm]{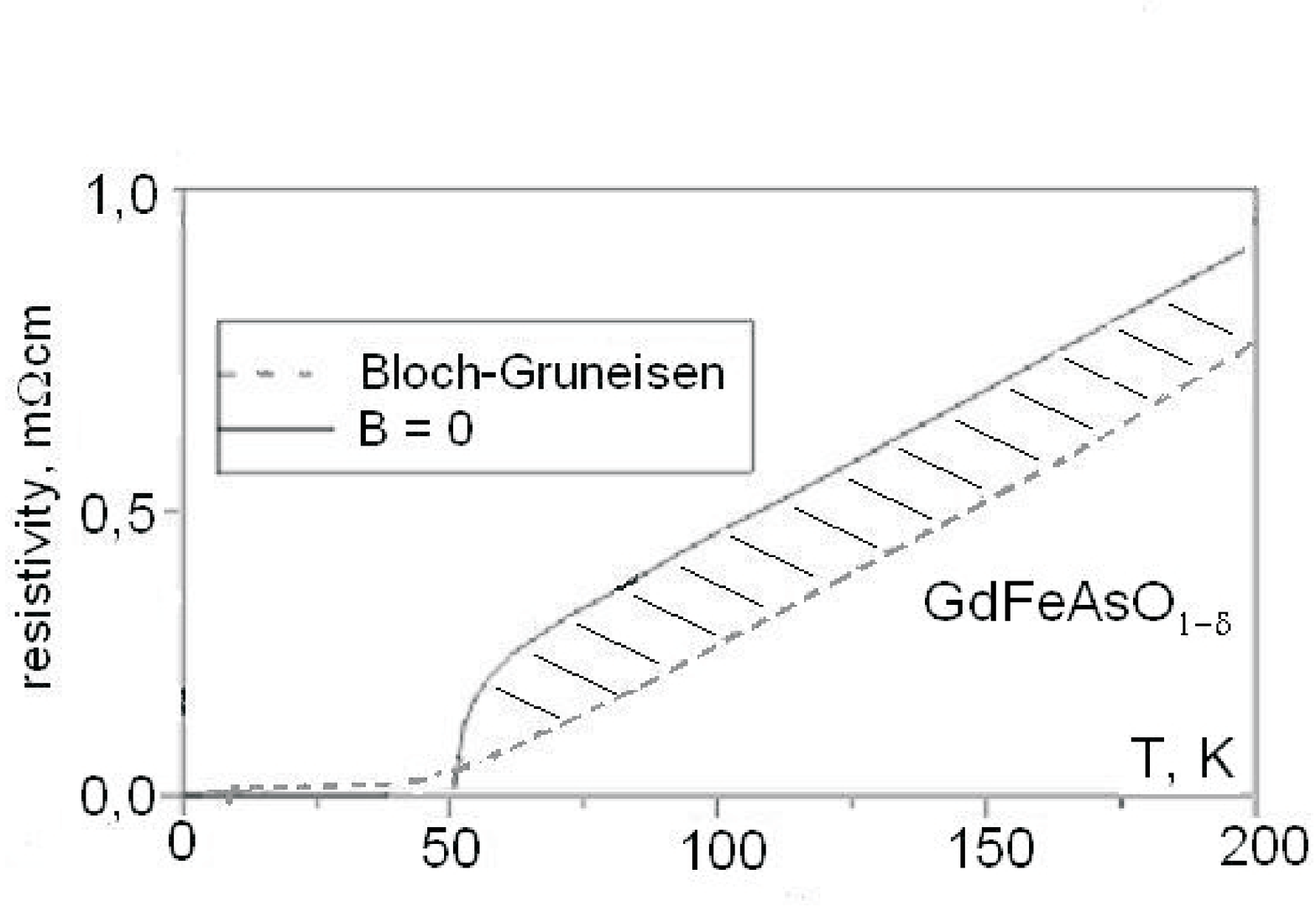}
\caption{Temperature dependence of resistivity for
$GdAsFeO_{1-\delta}$ sample (data are close to
\cite{GdFeAsO}).}\label{disp1}
\end{figure}
$$ \rho_{tot}(T) = \rho_{ph}(T) + \rho_m(T), \eqno(1) $$ where
phonon contribution is usually considered as corresponding to well
known Bloch-Gruneisen expression (see, e.g. \cite{Zim}) $$
 \rho_{ph}(T) = \rho_1
(\frac{T}{\Theta_D})^5 \int\limits_0^{\Theta_D/T}
\frac{x^5dx}{(exp(x) - 1)(1 - exp(-x))} \quad , \eqno(2) $$ and
magnetic contribution $\rho_m(T)$ in paramagnetic region as (see,
e.g. \cite{Vons}) $$ \rho_m(T)
=\frac{(m/m_0)^2NG^2s(s+1)}{(\pi/3)^{1/3}e^2h^3}n^{-2/3}. \eqno(3)
$$ The scaling parameter $\rho_1$ in (2) is determined by the
slope of linear part of the $\rho(T)$ dependence, and parameters
in (3) are usual (see, e.g. \cite{Vons}): $G$ is the coupling
constant, $s$ is the spin of fluctuating localized magnetic
moments (LMM), $N$ is the effective number of these LMM in unit
volume, $m$ and $n$ are the effective mass and density of mobile
charge carriers, respectively.

It is essential to note here that in (2), in addition to high-$T$
linear region there exists an intermediate-$T$ linear region
($0.22 \le T/\Theta_D \le 0.43$) characteristic for most metals
(including magnetic ones) \cite{Zim}. Namely, the latter, linear
in $T$ (not proportional (!) to $T$), region was considered in our
earlier work \cite{M1} as corresponding to the normal state of the
cuprates (see, also \cite{MPR}).

The above decomposition procedure (see (1)) performed for
$GdAsFeO_{1 - \delta}$ sample (cf. with \cite{GdFeAsO}) is
presented in Fig.1. Note, that here $GdAsFeO_{1 - \delta}$ is
taken only as an example: such almost linear in $T$ dependences
are also observed in other superconducting samples with optimal
doping in this family of Fe-based compounds (see, e.g. resistivity
data for $SmO_{1-x}F_xFeAs$ in \cite{SmOFFeAs, Ding}).

In Fig.1, solid line corresponds to experimental $\rho(T)$
dependence and dashed curve is a Bloch-Gruneisen (BG) one (see
(2)). The data for BG-curve are fitted by us with $\Theta_D
\approx 320 K$ which value is close to that estimated from
specific heat data for such compounds (see, e.g. \cite{SH}).

Then, the magnetic contribution to resistivity defined as
$\rho_m(T)= \rho_{tot}(T) - \rho_{ph}(T)$ (see (1)) corresponds to
the shadow region in Fig.1. Such suggestion is based on the fact
that parent compounds $ReOFeAs$ are magnetic in nature, and with
doping the AF spin fluctuations persist in the system providing an
effective channel for scattering of mobile charge carriers in the
normal state. (Note, such approach was before used by us for
detailed analysis of the resistivity in cuprates since they are
also of magnetic nature: parent compounds $YBa_2Cu_3O_6$ and
$La_2CuO_4$ for SC cuprates as known are antiferromagnetic
insulators.)

From Fig.1 it is seen that there are two distinct regions in this
$\rho_m(T)$ dependence. Well in the normal state ($T >T_c$) the
magnitude of $\rho_m$ is high enough and nearly independent of
temperature. Such behaviour corresponds to spin-disorder
scattering of mobile charge carriers in paramagnetic region for
magnetic metals and alloys and, in conventional theory,
resistivity due to such scattering can be expressed via (3) (cf.
with \cite{Vons}).

On the other hand, with decreasing temperature the magnitude of
$\rho_m(T)$ decreases, and moreover, in the point of intersection
of experimental $\rho(T)$ and BG curves, magnetic contribution to
resistivity $\rho_m(T)$ disappears, so that at $T$ below this
intersection point the resistivity drop occurs only due to phonon
contribution to resistivity. This disappearance of magnetic
contribution at low temperatures is, in fact, the evidence for
appearance of some magnetic ordering in the system, e.g. in the
form of modulated magnetic structure (cf. with \cite{Izyu}) say,
spin density wave (SDW) formed in the system near above $T_c$. So,
two different magnetic phases exist in this system at low and high
temperatures, respectively, and thus, decrease of $\rho_m(T)$ down
to zero with decreasing temperature can be considered as indirect
evidence for magnetic (AF SDW) phase transition in this system
before the SC transition.
\begin{figure}
\includegraphics[width=10.5 cm]{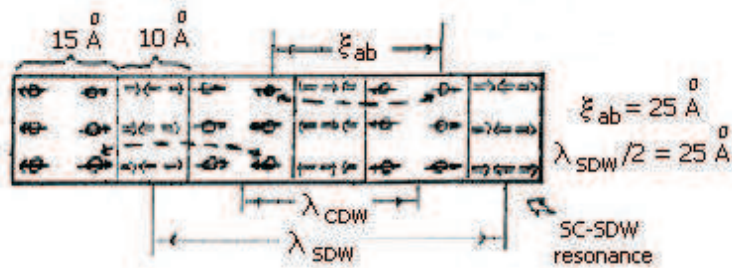}
\caption{The stripe structure in $CuO_2$-plane (scheme)(see, also
\cite{M06}).} \label{disp1}
\end{figure}
In other words, resistivity drop due to SC transition in these
compounds begins only from the point of intersection of
experimental $\rho(T)$ and BG curves, and part of resistivity drop
above this intersection point is attributed to magnetic (AF SDW)
phase transition, i.e. the resistivity drop above the intersection
point is essentially the normal state rather than SC one as it is
usually considered.

Note that from the same analysis performed by us for cuprates
before (see, e.g. \cite{M1,MPR,M06}) it was suggested a stripe
picture in the $CuO_2$-plane (see Fig.2). According to numerical
simulation in the Hubbard model on a two-dimensional square
lattice (for details, see \cite{Kato}), the arrangement of each
spin stripe is antiferromagnetic but two adjacent spin stripes are
alternate to one other so that in the $CuO_2$-plane it is formed
SDW with wavelength  $\lambda_{SDW} = 50 \AA$. The width of
stripes in Fig.2 corresponds to that obtained in \cite{Bia}. This
SDW, because of its incommensurability, is accompanied by CDW with
wavelength equal to one half of  that for SDW: $\lambda_{CDW} =
\lambda_{SDW} / 2 = 25\AA$ . Then, as it was obtained by us before
\cite{M1}, the in-plane coherence length  $\xi_{ab} = 25 \AA$, so
that there is SC-SDW resonance in the system (cf. with Fig.1,2).

As it's seen from Fig.2, for given stripe structure, electrons in
charge stripes (arrows with circles) are oriented by spin stripes
(double arrows) in a manner characteristic for the Cooper pairing
in the s-wave BSC theory. (Note that same conclusion on the s-wave
SC pairing symmetry follows from Fig.1, where SC transition starts
only from point at phonon GB-curve as in conventional
low-temperature superconductors (LTSC)).

\section  {THEORETICAL MODEL}

    The above picture is consistent with theory of interplay between magnetism and
superconductivity in itinerant electron systems \cite{Mach}. In
such systems with interplay between SC and magnetism, with
decreasing temparature, an itinerant SDW gap with magnitude
$\Sigma$ can appear at the Fermi surface only before the SC gap
($\Delta$) (see Fig.3), i.e. in the normal state. This SDW gap is
highly anisotropic since it is only formed at symmetric parts of
the Fermi surface \cite{Mach}( see, inset in Fig.3). Its magnitude
$\Sigma$ being unusually large for the SC gap , well conforms to
that for the SDW gap because of inequality $\Delta < \Sigma$ which
is peculiar for the coexistence phase in that model (Fig.3). In
such a case, the temperature $T_c^{onset}$ (see, Fig.3) can be
related to the appearance of the SC gap which begin develop at the
Fermi surface only when the transition of the system to a
magnetically-ordered (SDW) state is over.

Then, interrelations : $ T_c^{onset} \le T_{SDW}^{onset}$ and
$T_c^{onset} = T_m^{order}$ can be a natural consequence of the
equality of magnetic ordering energy $\varepsilon_m^{order}$ and
the condensation pair one $\varepsilon_c^{pair}$ considered as
characteristic for such itinerant electron system with an
interplay between SC and magnetism. (Note here that namely such a
picture was reported from elastic neutron scattering experiments
(see, e.g. \cite{Suzu}) when in La -based cuprate both the SC and
static magnetic order appear at the same temperature).
\begin{figure}
\includegraphics[width=9.5 cm]{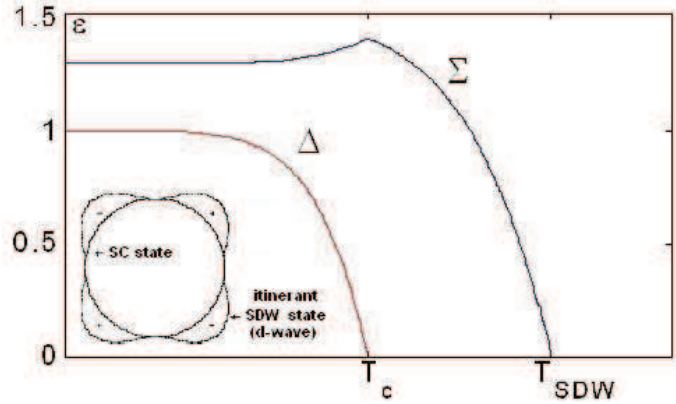}
\caption{Temperature dependence of SC and SDW order parameters
(scheme). Inset: symmetry of SDW and SC states (for more details,
see, e.g. \cite{M07}).}\label{disp1}
\end{figure}

Then, temperature and field behavior of such a system is
essentially determined also by amplitude fluctuations of local
spin density (FLSD)\cite{Moriya}: only slight effect of applied
magnetic field (paramagnetic state) because of temperature-induced
LMM regime of FLSD (spin disorder) above some temperature $T^*(H)$
in contrast to the enhancement of spin fluctuations by external
magnetic field below $T^*(H)$  characteristic for AF systems (for
references, see, e.g. \cite{MPR}).

Such a theory is essentially based on the model with partial
dielectrization of conduction electron energy spectrum (see, e.g.
\cite{DCK,PHTSC}). So, for systems with both (e-h)- and
(e-e)-pairing in which the SC and dielectric (DE) order parameters
can coexist with one other, the Hamiltonian can be written as sum
of three terms $$ H = H_i + H_{i,j} + H_{1,2}, \eqno(4) $$ where
$i$ and $j$ are the band indices ($i$, $j$ = 1,2) and $H_i$ and
$H_{i,j}$ describe intraband and interband interactions,
respectively, while third term corresponds to the interband
interactrion between electrons and static deformation $u(x)$ of
the crystal lattice with coupling constant $g_{12}^{eph}$.

By using the method of temperature Green function \cite{DCK} the
matrix equation for the given system can be expressed as $$
A(\Delta_{i,j}, \Sigma, \zeta_i,\omega_n)\times B(G_{i,j},
F_{i,j})= colon(1,0,0,0), \eqno(5) $$ where $G_{i,j}$, $F_{i.j}$
are the Green functions, $\Delta_{i,j} = g_{i,j}F_{i,j}(x,x)$,
$\Sigma = \tilde g_{21}G_{21}(x,x)$; $\zeta_i(p) = \mu \pm (p^2/2m
- \varepsilon_F)$, $\mu$ is the shift of the Fermi level in each
band due to doping; $\omega_n = \pi T(2n +1)$, $n$ is the integer
number; $g_{11}$, $g_{22} < 0$ ((e-e)-attraction) are the
constants of intraband interaction in the first term of (4) and
responsible for electron-phonon (e-ph) and for Coulomb interaction
(weakened due to the logarithmic factor
$ln(\varepsilon_F/\omega_D))$; $g_{21} (< 0)$ is the analogous
constant of interband interaction in the second term of (4);
$\tilde g_{21} = g_{21} + 2(g_{21}^{eph})^2/\omega_D > 0$ is the
constant responsible for (e-h)-pairing. The determinant for the
system (5) can be written in the form \cite{DCK} $$ Det =
(\omega_n^2 + \omega_+^2)(\omega_n^2+\omega_-^2), \eqno(6) $$
where $\omega_{\pm}$ is the energy of elementary excitations in
coexistent (SC + DE)-phase  (see Fig.4) $$ \omega_{\pm}^2 =
(\varepsilon \pm \tilde \mu)^2 \quad \quad \varepsilon^2 = \zeta^2
+ \tilde \Sigma^2. \eqno(7)$$

The expression (7) in fact determines the dispersion relation for
the system so that the electron energy spectrum in the
(SC+DE)-phase of such systems is characterized by presence of two
energy gaps (with different symmetry, in general case (cf. with
\cite{K70})) at the Fermi surface: DE-gap with magnitude $\Sigma
(T)$ due to (e-h)-pairing and SC-gap with magnitude $\Delta(T)$
due to (e-e)-pairing (see, Fig.4).

From above theory it follows that in the systems under
consideration with decreasing temperature firstly a structural
($\it dielectric$) transition occurs at $T = T^*$ ($T^*= T_{SDW}$
in our treatment) \cite {DCK,PHTSC,GEK}. In result, a modulated
lattice structure is formed in the system, and  at $T = T^*
(T_{SDW})$ the DE(SDW)-gap $\Sigma$ begins to develop at
symmetrical parts of the Fermi surface. With further decreasing
$T$ the magnitude of the DE(SDW)-gap $\Sigma$ increases, then  at
$T = T_c$ the SC transition occurs and the SC gap $\Delta$ begins
develop at the Fermi surface with decreasing temperature while the
order parameter  $\Sigma$ becomes to decrease gradually
approaching its zero-temperature value  $\Sigma (0)$ (see, Fig.3).
\begin{figure}
\includegraphics[width=10.5 cm]{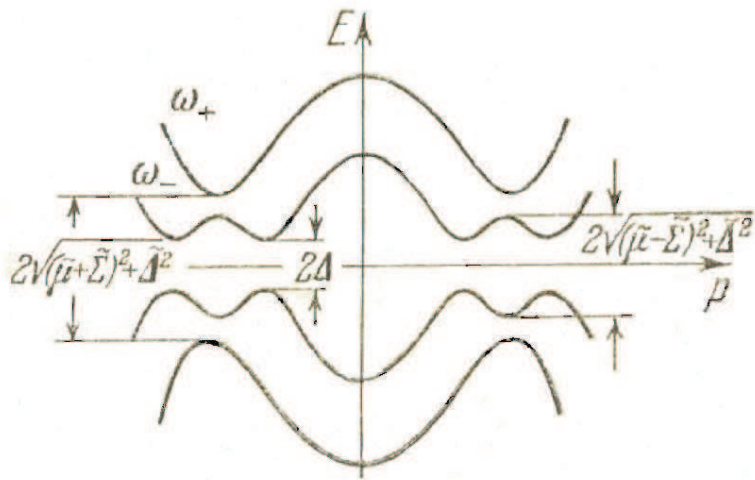}
\caption{Schematic electron energy spectra of coexistence phase of
SC- and DE-pairings (see \cite{PHTSC}).} \label{disp1}
\end{figure}

The solution of (5) corresponding to the SC-gap in the (SC +
DE)-phase (coexistence of (e-e)- and (e-h)-pairing) is given by $$
\tilde \Delta = 4 \tilde n^2 exp(-\tilde n\beta^*/(\Sigma_0-\tilde
n))/\Sigma_0, \eqno(8) $$ $$ \beta^* = \beta_0/(1-\beta_0
\lambda_{21}\Sigma_0(\Sigma_0 - 2\tilde n)/(\Sigma_0 -\tilde
n)^2), \eqno(9) $$ where $\Sigma_0$ is the DE-gap in absence of
doping; $\tilde n = \delta n / 4N(0)$, $\delta n$ is the
difference of electron and hole densities, $N_0$ is the density of
states near  the Fermi  level; $\beta_0 = ln(\Sigma_0/\Delta_0) >
0$,  $\Delta_0$ is the SC gap in absence of DE-pairing;
$\lambda_{21} = g_{21}N(0)$ \cite {PHTSC, DCK}.

On the other hand, for the DE-gap magnitude $\Sigma$ it can be
used the expression from \cite{KK} $$ \Sigma = 2 \omega_0 exp(2\pi
v_F/ e^2ln(\kappa_D^2/2p_0^2)), \eqno(10) $$ which, in fact,
describes the insulator with thermodynamics similar to that of
superconductor.

\section{Discussion}

So, from above it follows that there should be two gaps (SC and
SDW) with different nature in electron energy spectra of Fe-based
superconductors at low temperature. Of course,  fact  of existence
of two (SC and SDW) energy gaps (two (SC and SDW) order
parameters) in the system should be taken into account at
treatment of experiments, first of all, dealing with threshold
phenomena like infrared absorption, nuclear spin relaxation,
hypersonic absorption, tunneling experiments etc. in the
(SC+DE(SDW in nature))-phase of such systems \cite{PHTSC,KopUFN}.
It can be noted that at present, ARPES experiments should be also
added to this list.

Since there are rather a few data for iron-based superconductors,
and moreover, they are mainly obtained with polycrystalline
samples then below there are also presented experimental data for
cuprates, picture in which is believed to be essentially similar
to that in iron-based compounds.

\subsection{Manifestation of two energy gaps (two order parameters) in experiment}

{\bf (AR) PES}. Indeed, the evidence for two different energy gaps
with different energy scale was recently observed in photoemission
spectroscopy (PES) measurements in F-doped $LnOFeAs$
polycrystalline sample \cite{Ishida,Liu}. However, since (AR) PES
technique measures below $T_c$ only the combined energy gap: $E =
\sqrt{\Delta^2 + \Sigma^2} $ rather than each of them separately,
then it is necessary to take this fact into account under treament
of measurements in superconducting state ((SC+SDW)-phase of
Fe-based system), (cf. with that in cuprates, \cite{M07}).

{\bf Neutron scattering}. From measurements with using of the
neutron scattering technique up to now there was no observed
evidences for SDW in doped Fe-based compounds in both normal and
SC state, for both cases with and without of applied magnetic
field \cite{Qiu}.

However,  in cuprates, in neutron diffraction experiments
\cite{Lake}, it was obtained the evidence for antiferromagnetism
of the vortex core in resistive state which in the underdoped
single crystal of $La_{2-x}Sr_xCuO_4$ (x=0.1) is static in
character (elastic neutron scattering) while in the optimally
doped sample $La_{2-x}Sr_xCuO_4$ (x=0.163) enhancement of low
frequency spin fluctuations in applied magnetic field was only
observed (inelastic neutron scattering), and the vortex core
antiferromagnetism was treated as field-induced for both samples.
Moreover, the AF regions were also indicated in the space between
vortices. Then, antiferromagnetic (SDW) ordering develops in LSCO
just below the zero-field SC transition temperature $T_c(H=0)$ and
increases with decreasing temperature and increasing field.
However, this ordering in LSCO was considered in \cite{Lake} only
as field-induced though the zero-field magnetic response (when
there are no SC vortices in the system) is also presented in the
SC state, and the onset temperature of so defined AF (SDW) phase
transition in zero- field case coincides, in fact, with that of
zero-field SC transition: $T_m(H=0)=T_c(H=0)$ (cf. with Fig.5).

On the other hand, in work \cite{Khay} where a zero-field elastic
magnetic signal was observed, increasing with decreasing
temperature and saturating at $T\to 0$, the behavior was noted as
correspondent to the classical mean-field theory. Moreover, the
same temperature behavior for the incommensurate magnetic signal
was also observed in external magnetic field which fact was there
treated as enhancement (rather than induction) of long-range
magnetic order by magnetic field. Then, there was obtained a
linear dependence of saturated at $T \to 0$ magnetic response on
the magnetic field value characteristic for the vortex state.
\begin{figure}
\includegraphics[width=9.5 cm]{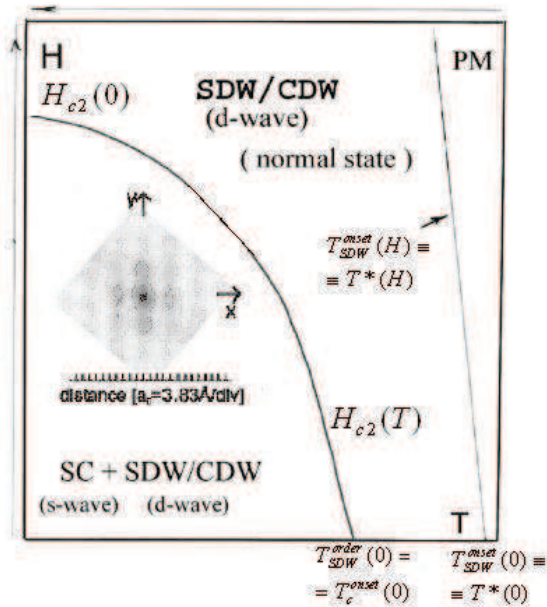}
\caption{Magnetic phase $H - T$ diagram for itinerant electron
system with interplay between SC and magnetism. Inset: The
in-plane checkerboard pattern in the vortex state of BSCCO single
crystal from magnetic-field STM measurements (after \cite{Hoff}).
}\label{disp2}
\end{figure}
However, the onset temperature of the observed AF (SDW) phase
transition was estimated as field-independent.

Note here, that nearly the same picture of modulated low-frequency
spin fluctuations was obtained in YBCO single crystals from
neutron scattering experiments in both the normal and
superconducting states (see, e.g., \cite{Mook}). Moreover, a small
magnetic signal was observed below $T \approx$ 200 K which then
increases below $T_c(H=0)$.

{\bf Mossbauer spectroscopy}. From $^{57}Fe $ Mossbauer
spectroscopy measurements  on La-family of  Fe-based
superconductors there was no observed also any evidence for SDW
order in the whole temperature range of interest (see, e.g.
\cite{Hosono2}). However, in \cite{Nowik} on the basis of the same
measurements as well as magnetic ones  in $SmFeAsO_{0.85}$
superconductor it was obtained an evidence for coexistence of SC
and magnetism in the FeAs-layer (cf. with Fig.2 in present paper).
Then, from the Mossbauer spectra in the normal state it was
obtained a well-defined quadrupole doublet while in the SC state a
magnetic sextet was developed. The possibility of  these
observations could be provided due to the coupling between the Sm
and Fe moments in the system which is absent in LaOFeFAs
superconductors.

{\bf $\mu$SR}. From muon spin relaxation measurements (which a
time window is usually in the range between $10^{-6}$  and
$10^{-9}$ s \cite{Drew}) in $LaO_{0.97}F_{0.03}FeAs$ it was
obtained an evidence for possible incommensurate or stripe
magnetism (cf. with Fig.2 in present paper): Bessel-function line
shape of time spectra of zero-field $\mu SR$ characteristic for
such magnetism was observed for doped compounds \cite{MuSR}.

{\bf Tunneling}. In the point-contact spectroscopy measurements
there was obtained a peak-dip-hump structure in tunneling spectra
\cite{Lei} which is reminiscent to the case of cuprates where such
structure was obtained e.g., in STM measurements \cite{Krasnov}.
There, from such measurements with single crystal Bi-2212 it was
obtained the peak-dip-hump structure considered as evidence for
two energy gaps in the electron energy spectra and a pseudogap
appears to be in the role of the DE-gap (note that according to
our treatment, the PG is of SDW/CDW nature) (for more details, see
\cite{M07}). Moreover, while in \cite{Krasnov} the SC-gap with
increasing $T$ disappears at $T = Tc$ ($\Delta  = 0$), the DE-gap
(PG(SDW)) magnitude is not zero and $\Sigma(T)$ vanishes only well
in the normal state.

On the other hand, from magnetic-field STM experiments with
$Bi_2Sr_2CaCu_2O_{8+\delta}$ single crystal \cite{Hoff}, on the
basis of detailed comparison of STM pattern of the sample surface
in high magnetic field with that in zero-field case, a
checkerboard pattern was visualized inside and outside the vortex
core (see insert in Fig. 5). There, it was concluded also about
the field-induced nature of the periodicity in the measured local
density of electron states (LDOS). Though the STM method is
insensitive to the magnetism, it was proposed by them that the
incommensurability of the in-plane LDOS modulation observed below
$T_c(H=0)$ corresponds to the SDW/CDW picture (stripe structure
with alternating spin and charge stripes (for review, see
\cite{Bia})) with the wavelength $\lambda_{CDW} = \lambda_{SDW}/2$
characteristic for general theory of SDW systems (see, e.g.,
\cite{TUG}) which interrelation between spin and charge stripes
was actually obtained there.

Note here, that in addition, in another low temperature
magnetic-field STM measurements in cuprates the evidence for the
pseudogap with $d$-wave symmetry in the vortex core (the same as
in the normal state) was observed so that SDW nature of the
pseudogap was proposed (for references, see \cite{M04}).

{\bf NMR}. NMR experiments performed with novel Fe-based
superconductors can be also considered as consistent with  two
energy gap (SC and SDW) picture in above theoretical model (see,
Sec.3). So, in $^{75}As$ NMR measurements performed on
$LaO_{0.9}F_{0.1}FeAs$ superconductor \cite{NMR} it was obtained
the evidence for pseudo-spin gap in the normal state which gives
rise to a suppression of  magnetic shift $K_s$ arising via a
hyperfine coupling to electron spins and spin lattice relaxation
rate $T_1^{-1}$ for temperatures below 300 K, moreover, decreasing
of magnetic shift in the SC state is considered in \cite{NMR} as
suggestive for spin-singlet pairing. The temperature dependent
Knight shift in F-doped $LaOFeAs$ was also concluded to be similar
to pseudogap (SDW-gap in present treatment) behaviour in cuprates.

In this connection, it should be noted recently presented results
of spatially resolved high-field NMR studies of optimally doped
YBCO (for references, see \cite{M07}) where indication was also
obtained for the presence of correlated antiferromagnetic
fluctuations in the vortex core. The method used the similarity of
a spatial distribution of the internal magnetic field in the
vortex lattice and the NMR spectrum for any nucleus with small
intrinsic broadening (e.g., for $^{17}O$ in YBCO).

It should also be noted that above proposal was supported by
magnetic-field NMR measurements in cuprates, in which the absence
of any shift of the pseudogap (SDW, in our treatment) onset
temperature $T^*$ up to high enough magnetic field was observed
while the SC transition temperature $T_c(H)$ was essentially
shifted downwards to zero \cite{Gorn}. This fact was considered
there as evidence for the relatively large energy scale for the
pseudogap mechanism which scale is rather characteristic for the
density waves (e.g., SDW/CDW state).

\subsection{Static vs. quasi-static additional (magnetic) order
in superconducting Fe-based compounds}

Note, that though in cuprates such magnetic phase transition has
no been directly seen before but its possibility was discussed
anywhere and some evidences was obtained (for details, see
\cite{M04}). Moreover, as it was noted in a number of works, it's
direct observation by the normal neutron scattering can appear to
be problematic because of specificity of the low-temperature
quasi-static magnetic ordering. The same situation can be realized
in Fe-based compounds as well.

So, from above it can be concluded that in these compounds below
$T_c(H=0)$ a periodic spatial modulation of magnetization
considered as static (elastic peaks) or as fluctuating (inelastic
signal) one is observed. Such a behavior can be regarded as
evidence for the AF SDW ordering which is dynamical in nature, in
general case. In this sense, the low energy (i.e., dynamic)
incommensurate spin excitations in the system can correspond to
the dynamic formation of stripes and moreover, in cuprates,
dynamic nature of stripes is even stronger in LSCO ($x \approx
0.125$) though stripes are frequently (and erroneously) considered
as static below $T = 20 K$ based on elastic neutron scattering
data. In addition, it was demonstrated in \cite{Hunt} that static
stripes are not really static even at 350 mK. Thus, even "static"
ordering (stripe structure) cannot be regarded as really static
but only depending on the timescale of the local probe (cf. with
\cite{M04}). In other words, the question 'is the observed (SDW)
order static or fluctuating one' is determined by the relation
between characteristic frequency of spin fluctuations and
timescale of concrete local probe. So, according to \cite{Hunt},
the critical temperature to observe the spin ordering $T_{spin}$
is lower for experimental probes with slower frequency scales,
i.e., $T_{spin} \sim 50 K$ for elastic neutron scattering ($\sim
10^{11} Hz$), $T_{spin} =30 K$ for $\mu SR$ ($\sim 10^7 Hz$) and
$T_{spin} = \sim 10-30 K$ for NQR ($\sim 10^6 Hz$). In addition,
the spin fluctuation frequency appears to be dependent on
temperature so that in some temperature regions in cuprates it was
observed a slowing down effect which leads to wipeout of the NMR
signal and, in result, the SDW ordering (stripe structure) appears
to be undetectable with the given local probe. In this sense, such
a problem is in fact absent for the conduction electron scattering
which provides the electrical resistivity of metals. Indeed, in
resistivity measurements, due to short mobile charge carrier
relaxation time, a magnetic structure is sampled on much shorter
time scale as compared with many other techniques such as neutron
scattering, Mossbauer effect (cf. with neutron and Mossbauer
spectroscopy results in Fe-based superconductors in Sec.4A, see
above), etc. (see, also \cite{M04} and references therein).

Then, note also that during the study of cuprates it was
introduced a concept of so-called 'hidden' order in the cuprates
(for references, see, e.g. \cite{M04}). Such an order is
attributed to $d$ -density wave (DDW) order. However, in their
statement the type of this DDW order is not concrete but it is
only considered as competing (not vital) order for SC one,
moreover it is considered as corresponding to the
superconductivity with $d_{x^2 - y^2}$ -wave pairing symmetry. As
follows from above this concept can be described in terms of the
(spin) density wave (S) (DW) (and (charge) density wave (C) (DW)
with $\lambda_{CDW}= \lambda_{SDW} / 2$) state with a $d_{x^2 -
y^2}$ -wave symmetry (see above). The effects of these density
waves (DW) are well known and the 'hiddenness' of these (DW) in
the cuprates can be of natural consequence of the dynamical nature
of these (S) and (C) (DW) so that only fast and local probe
(including resistivity measurements) permits to detect these (DW)
concretely.

\subsection {Phase diagram of  Fe-based superconductors}.

From above theory it follows that doping dependence of magnitude
of  SC gap ($\Delta$) and DE (SDW) one ($\Sigma$) in such a system
should be quite different. The superconducting gap (8) (as well as
critical temperature $T_c$) has a maximum as a function of  the
carrier density. This maximum appears due to the competition
between increase in the density of states at the edges of the
dielectric gap (singularity) as result of (e-h)-pairing and
decrease of energy interval with electrons near the Fermi level as
result of the DE (SDW) gap formation \cite{DCK,PHTSC}. Such
conclusion is also consistent with expression obtained in
\cite{K70} $$ T_c = 1.14 (\omega^2)^{1/2}(\mu/\Sigma)
exp{(-(\mu/\Sigma)/\lambda N(0))}, \eqno(11) $$ where $\lambda$ is
the effective constant of (e-ph)-interaction, $\omega$ is the
average value of phonon frequency. Moreover, maximum value of
temperature of superconducting transition can be essentially
higher as compared with $T_{c0}$ in absence of dielectric (SDW)
(e-h)-pairing \cite{DCK,PHTSC} $$ (T_c/T_{c0})_{max} = 4 exp
\beta_0/e\beta_0, \eqno(12) $$

As for DE(SDW) gap $\Sigma$ then it (and SDW onset temperature
$T^*$) is decreasing function of doping (see (10)). So, as a
result, these two dependencies in fact determine the phase diagram
of the Fe-based system.

In this connection, there can be interesting the recent results of
electronic Raman scattering experiments \cite{Tacon} in which
indication to two energy scales in the SC state of cuprates are
presented. First, characteristic energy (corresponding to
magnitude of the SC-gap $\Delta$, and thus to that of $T_c$)
measured in a nodal region of momentum space has a maximum as a
function of doping (cf. with (8) and (11)). Such
$\Delta(p)$-dependence with maximum is characteristic for the SC
order parameter with s-wave symmetry in the model of
\cite{DCK,PHTSC}.  In contrast, second characteristic energy
(corresponding to magnitude of the DE(SDW)-gap  $\Sigma$ and thus
$T^*$) measured in antinodal region appears to be decreasing
function of doping, in agreement with theory \cite{KK}. Such a
picture is consistent also with our model for symmetry of the two
order parameters in cuprates: s-wave for SC one and d-wave for
DE(SDW) one (see insert in Fig. 3)(cf. with \cite{M06,Klemm}).

\subsection{Magnetic phase $H - T$ diagram; upper critical field $H_{c2}(0)$}

The above decomposition procedure (see (1)) permits to obtain the
magnetic phase $H-T$ diagram in the whole temperature range of
interest (see Fig.5) from resistivity data in a magnetic field by
a correct way. In contrast to the problem with so-called 'upward
curvature' of $H_{c2}(T)$ curve near $T_c(H = 0)$ discussed in
literature, e.g. in 'mid-point' or 'zero-resistance' methods, the
decomposition procedure (see (1)) provides correct data for this
diagram from points of intersection of experimental $\rho(T)$
dependencies in applied magnetic field with BG curve, and so
formed $H_{c2}(T)$ dependence appears to be linear in $T$ near
$T_c (H = 0)$ as in BCS and GL theories (see, e.g. \cite{M1,MPR}).
The data to form a $T^*(H)$ dependence for SDW can be taken from
characteristic points at experimental $\rho(T)$ dependencies in a
magnetic field corresponding to beginning of deviation from
linearity in $T$ with decreasing temperature (for details see, e.g
\cite{M1,Ito,MPR}). (Note that according to above (Sec. 4B), the
magnetic phase $H-T$ diagram for compounds under study (see Fig.5)
should be considered as dynamic in nature).

In this magnetic phase $H-T$ diagram the results discussed above
can be accounted for by a natural way. The point in the
(SC+SDW)-state (antiferromagnetic (SDW) order in the SC state) at
the ($H, T$) plane can be reached by two different ways. First, it
can be reached under cooling in a constant external magnetic
field. But, if one will reach this point by increasing the
external magnetic field at given temperature then
antiferromagnetic (SDW) order will be looked as 'field-induced'
though it has only emerged in the core of vortices since, and as
follows from above, the SDW/CDW state can be regarded as
underlying order for the SC one in the cuprates and Fe-based SC
compounds.

\subsection{On ferromagnetism and superconductivity in novel superconductors}

It is sometimes considered as surprizing that strong magnetic
element $Fe$ normally giving rise to magnetic moments, and in many
cases forming a long range ferromagnetic order, and being thus
detrimental to the superconductivity with singlet pairing play an
important role in superconductivity. However, the problem of
coexistence of ferromagnetism and superconductivity was else
discussed by Ginzburg in his earlier work about ferromagnetic
superconductors \cite{Ginz}. There it was demonstrated that
ferromagnetism and superconductivity can coexist with one other if
the internal ferromagnetic molecular field is weaker compared with
zero-temperature thermodynamic critical field $H_{c2}(0)$ of
superconductivity.  In other words, under conventional conditions,
ferromagnetic sample can enter the SC state only in exclusive
cases, when its sponaneous magnetization $M_0$ (at $T = 0$) is
very low ($B_0 = 4\pi M_0 \leq 1 T$), while upper critical field
should be high enough ($\ge 1 T$). Since the upper critical field
$H_{c2}(0)$ for conventional LTSC is between 0,26 T (for $Nb$) and
0,0026 T (for $Gd$) then it is not the case for LTSC but it
appears to be the case for novel Fe-based superconductors where
$H_{c2}$ was estimated as near above 60 T \cite{Hunte} (for
comparison, the value of $B_0$  for sample of pure $Fe$ is only
near 2,2 T). Note the results of \cite{SH} in which the
temperature of maximum in difference of electronic specific heat
coefficients $\gamma(H = 0) - \gamma(H = 9T)$ is consistent with
that of intersection point of experimental $\rho(T, H = 9T)$ curve
with BG one (cf. with Fig.1), i.e. with $T_c^{onset}$ at $H = 9T$,
from what it follows that the upper critical field $H_{c2}(0)$
estimated from the data in \cite{SH} according to BCS and GL
theories, appears to be near 20T. It is essential, however, these
values overestimate the real value of upper critical field for
$\vec H \| \vec c$ case due to polycrystalline structure of
samples, since $\vec c$-axes are randomly oriented into the
sample. But the latter value of $H_{c2}(0)$ will be, of course,
higher then 2,2 T (see, above).

\subsection{The nature of strong suppression of $T_c$ by
non-magnetic impurities in (SC+SDW) state}

Another issue which is described by the above theory (see Sec.3)
in a natural way is the suppression of $T_c$ by nonmagnetic
impurities which effect is not the case for LTSC. So, in the (SC +
DE)-coexistence phase the values of $\Sigma$ and $T^*$ being
influenced by scattering with nonmagnetic impurity, will be also,
in their turn, influent the values $\Delta$ and $T_c$ because of
strong correlation of later ones with $\Sigma$ and $T^*$ (=
$T_{SDW}$, in our treatment) (see, e.g. \cite{PHTSC}). This effect
of depairing of (e-h)-pairs by nonmagnetic impurities (in analogy
with depairing of (e-e)-pairs by magnetic impurities in LTSC)
determines both strong suppression of $T_c$ and decrease of the
DE-gap (PG, SDW/CDW-gap \cite{M1, Klemm}) with increasing the
impurity concentration which effect was observed in cuprate
systems,\cite{Pimen}. In contrast, in the case of magnetic
impurity, there should take place an increase of the magnitude of
DE-gap (PG, SDW/CDW-gap \cite{M1,Klemm}) (also observed in
\cite{Pimen}) due to polarization of impurity by the SDW, leading
to an exchange enhancement of the SDW, the effect well known for
AFM systems with itinerant electrons (for details, see
\cite{M07}).

\section{CONCLUSION}

From preliminary analysis performed above it can be concluded the
following. The doped Fe-based compounds are related to systems
with both (e-e)- and (e-h)-pairings due to electron-phonon and
Coulomb interactions, respectively. In such systems, with
decreasing temperature, firstly it occurs a dielectric (DE) phase
transition (and structural one) at $T = T^*$ (due to
(e-h)-pairing) and only then SC one (due to (e-e)-pairing) at $T =
T_c$ so that below $T_c$ the system enters the coexistence (SC
+DE) phase. The thermodynamics of dielectric transition is the
same as for superconductor. During this process, the dielectric
gap is formed only at the part (s) of the Fermi surface. High
$T_c$ of  the SC transition in such system is reached since
(e-e)-pairing occurs at the background of high density of states
(singularity) arising in energy region near dielectric-gap edges
due to removing of electronic states from the energy region of the
dielectric gap (already formed in the normal state). (It is well
known that in BCS-theory the SC transition temperature $T_c$ is
highly sensitive to the density of states value). Such conditions
arise when parameters of the system, changing with e.g. doping,
appear favourable for Fermi level to be located in the energy
region with high density of states (singularity) at the edges of
the dielectric gap. However, because of formation of the DE-gap at
the Fermi surface the energy region for (e-e)-pairing is decreased
what leads to decreasing of $T_c$. So, it should appear a maximum
in dependence of  $T_c$ (and hence of $\Delta$) as a function of
doping (optimal doping). In contrast, the DE-gap (and hence $T^*$)
is a decreasing function of doping. Thus, these two doping
dependences can be considered as forming a phase diagram of the
system.

Such a theory, being applied to itinerant electron system with
interplay between superconductivity and magnetism, leads to
formation of anisotropic SDW-gap (as dielectric gap) at
symmetrical parts of the Fermi surface well before SC transition.
This SDW in doped Fe-based compounds appears to be incommensurate
with lattice (in contrast to commensurate one in parent compounds
$ReOFeAs$) and it is accompanied by CDW with wavelength equal to
one half of that for SDW so that in FeAs-plane it should be formed
a stripe-like structure with alternating spin and charge stripes.
The SDW order parameter is dynamical in nature, and its detection
depends on timescale of  technique used. The magnetic field being
applied provides the periodicity (SDW/CDW structure) inside the
vortex core region at $T$ below $H_{c2}(T)$ curve at the magnetic
phase $H-T$ diagram, while above $T^*(H)$ curve at this diagram
the SDW state is absent at all.

Of course, the above analysis was performed on the basis of
available data for polycrystalline samples of doped Fe-based
compounds and more correct conclusions can be obtained only when
measurements on single-crystal samples of Fe-based superconductors
will appear, however, it seems that the general picture remains to
be the same.
\\

* Electronic address: mazov@ipm.sci-nnov.ru

\end{document}